\documentstyle[aps,prl,epsf,preprint]{revtex}
\def\fun#1#2{\lower3.6pt\vbox{\baselineskip0pt\lineskip.9pt
        \ialign{$\mathsurround=0pt#1\hfill##\hfil$\crcr#2\crcr\sim\crcr}}}
\def\mur{\mu_R}
\begin{document}

\title{\vskip-2.5truecm{\hfill \baselineskip 14pt {{
\small  CERN-TH/97-186\\
        \hfill hep-ph/9707537}}\vskip .1truecm} 
\vskip 0.1truecm {\bf A Simple Complete Model of Gauge-Mediated SUSY-Breaking
and  Dynamical Relaxation Mechanism for Solving the $\mu$-Problem}}
\author{{S. Dimopoulos}$^{(1),(2)}$, 
{G. Dvali}$^{(1)}$ {and R. Rattazzi}$^{(1)}$}
\address{$^{(1)}${\it Theory Division, CERN \\CH-1211 Geneva 23, Switzerland}}
\address{$^{(2)}${\it Physics Department, Stanford University\\
Stanford, CA 94305, USA}}
\maketitle


\begin{abstract}
\baselineskip 12pt

\end{abstract}

In this paper two things are done. First, we propose a simple model of
dynamical  gauge-mediated SUSY breaking.
This model incorporates a dynamical relaxation mechanism which
solves the $\mu$-problem with no
light fields beyond those of the MSSM.
In the second part of the paper we show how this mechanism  is generalized
and give two more examples in which it is incorporated.
\thispagestyle{empty}

\newpage
\pagestyle{plain}
\setcounter{page}{1}
\def\beq{\begin{equation}}
\def\eeq{\end{equation}}
\def\beqa{\begin{eqnarray}}
\def\eeqa{\end{eqnarray}}
\def\tr{{\rm tr}}
\def\x{{\bf x}}
\def\p{{\bf p}}
\def\k{{\bf k}}
\def\z{{\bf z}}
\baselineskip 20pt
\subsection{A Simple Model}
Gauge-mediated (GM) theories of supersymmetry breaking \cite{ancient} have
recently attracted a great deal of attention because they provide a natural 
solution to the supersymmetric flavour problem \cite{flavor} and have new 
phenomenological signatures. Since the original pioneering work\cite{DiNeSh},
there has been significant progress in building models which combine 
dynamical supersymmetry breaking with gauge mediation.
(see e.g. \cite{gd1} - \cite{gd8}).
In particular, new mechanisms  based on  quantum-modified moduli spaces
\cite{seiberg} have been explored \cite{iyit}. 
It has also been shown that a class of simple
strongly coupled gauge theories, which usually preserve
supersymmetry, do break it dynamically when 
combined with the stringy anomalous $U(1)$  \cite{dpbd}.

We have recently proposed a class of simple theories in which SUSY is 
unbroken in the global minimum, and yet the soft terms are dynamically
generated because we live in a metastable
SUSY-breaking plateau\cite{us}\footnote{One of our models
has a significant ideological overlap, including the problem of
tachyonic sfermions, with ref\cite{murayama}. An interesting model along 
similar lines was
recently proposed in \cite{luty}.}.  Unfortunately the most elegant of our 
theories had negative mass$^2$ scalars\cite{gianriccardo}.
In this paper we propose a simple class of realistic theories with such
 metastable vacua.
These theories
provide for a simple solution to the $\mu$-problem based on a dynamical 
relaxation mechanism \cite{dgp}; furthermore they do not require the 
introduction of any input mass parameters.
In the second part of the paper we show how to generalize this mechanism and
give two more examples incorporating it.

According to the general mechanism of ref.\cite{us}
our gauge group includes three factors
$SU(5)_S\otimes SU(5)_W\otimes G_B$. Here $SU(5)_W$ is the ordinary 
(weakly coupled)
grand unified group which accommodates quarks and leptons in three
families of $(1,10,1) + (1,\bar 5,1)$ representations.
For our discussion it is not
essential that $SU(5)_W$ be realized as a gauge symmetry of the theory
above the GUT scale. In particular, the $SU(3)_C\otimes SU(2)\times U(1)_Y$
subgroup would suffice. Nevertheless, for
convenience, we will classify fields by $SU(5)_W$ representations. $G_B$ is 
a weakly coupled `balancing group', whose role
is to stabilize the supersymmetry-breaking minimum
by a mechanism similar to that of 
Witten's `inverse hierarchy' scenario\cite{witten}.
This role can be played by
any asymptotically free gauge group with some vector-like matter content. For
 definiteness we will take $G_B=SU(2)_B$ with a single flavour
$\psi, \bar{\psi}$. Finally, $SU(5)_S$ is a strongly coupled group with
five flavors $Q_i \bar Q^i$, where $i = 1,..5$ is a flavour index which
corresponds to $SU(5)_W$.
Thus the `quarks' of the strongly-coupled group are in the representation
 $(5,\bar 5, 1) + (\bar 5, 5,1)$.
We will denote the strong scale of $SU(5)_S$ by $\Lambda$.
To complete the matter content, we introduce four additional gauge singlets
$Y,I,S,~ {\rm and} ~N$. Their role will become clear later.

Our classical superpotential is
\begin{eqnarray}
W = \alpha Y{\rm Tr M} + I(\beta Y^2 - \gamma \psi\bar{\psi}) + 
S(\lambda_1 H\bar H + \lambda_2 N^2/2  - \lambda {\rm Tr}M)
\nonumber\\
+ {\rm standard~MSSM~(GUT)~ superpotential} \label{superpotential}
\end{eqnarray}
where $M_i^k = Q_i\bar Q^k$ are the `mesons' of the strongly coupled group
and $H,\bar H$ are the MSSM Higgs doublets. 

The role of the various terms in the superpotential can be briefly summarized
 as follows.
\begin{description}
\item [1)] After accounting for the non-perturbative dynamics, the first 
term produces
 a supersymmetry breaking plateau along which
$F_Y \sim \Lambda^2$ and $Y$ is undetermined.
\item [2)] The second term is the main improvement with respect to the models 
of  ref.\cite{us}. 
It relates the vacuum expectation value (VEV) of the singlet $Y$ to the 
VEV of $\psi\bar \psi$. Thus 
the balancing group is broken (`Higgsed') along the plateau and 
one-loop effects can create a stable minimum at $Y >> \sqrt{F_Y}$
via Witten's inverse hierarchy mechanism. In this minimum $Q, \bar Q$
automatically play the role of the usual messengers of  
supersymmetry breaking. Note that without the second term, the only
way to stabilize the SUSY breaking minimum is with a gauge-non-singlet
$Y$ as it was done in ref. \cite{us}. This would make it more
difficult to solve the
$\mu$ problem  and may also lead to negative 
sfermion mass${}^2$ \cite{gianriccardo}.
\item [3)] The last term solves the $\mu$ problem via the mechanism of
ref. \cite{dgp}: the $\mu$ and $B\mu$ terms are induced from the original
SUSY-breaking scale $F_Y$ as one- and two-loop
effects respectively.
\end{description}

Let us now discuss this model in more detail. First we study its 
classical vacuum manifold (the moduli
space). It includes a flat direction that can be parameterized by 
 $Y$. Along this direction the $Q, \bar Q$ states get masses $\sim Y$.
For $\beta = \gamma =0$ there would be  two other independent flat 
directions on the moduli space:
a trivial one, given by the free field $I$, and another, parameterized by
the invariant $\phi = \sqrt {2\psi \bar{\psi}}$. The latter is the only
flat direction allowed by the $SU(2)_B$ $D$-terms. Along it, $SU(2)_B$ is
completely Higgsed and its gauge superfields get a mass $=g_B\phi$
by eating up the three complex degrees of freedom from $\psi, \bar{\psi}$.
Thus, in the $\beta = \gamma = 0$ limit we would
have a moduli space of  complex dimension 3.
However, for $\beta, \gamma \neq 0$ two out of  three flat directions are 
lifted. This is because
the $F_I$ and $F_\phi$ flatness conditions force to have
\begin{equation}
 Y\sqrt{\beta} = \phi\sqrt{\gamma/2}\, ~{\rm and}~ \, I=0
\end{equation}
As a result $I$ and the combination $Y\sqrt{\beta} - \phi\sqrt{\gamma/2}$ 
 get  mass
$= 2\sqrt{\beta^2 + \beta\gamma/2}\langle Y \rangle$. Notice that we define
 $\beta$ and $\gamma$ to be positive and real.
The other
superposition remains flat and we parameterize it by
$X=Y\sqrt{1+2\beta/\gamma}$. 
Thus, perturbatively the massless degrees of freedom along this branch of the
moduli space 
are: the $SU(5)_S$ vector  supermultiplet, the chiral superfields $X$, $S$ and 
$N$, plus the ordinary MSSM spectrum (quarks, leptons
etc.). Integrating out the heavy modes,
the perturbative effective low energy superpotential is
\begin{equation}
W =  S(\lambda_1 H\bar H + \lambda_2 N^2/2)
+ {\rm standard~MSSM~(GUT)~ superpotential}
\end{equation}
We see that $X$ simply dropped out of the superpotential. The K\"ahler 
potential 
for $X$ is determined by the wave functions of the the fields in the original 
theory as discussed in ref. \cite{us}
\begin{equation}
K(X,S)=\left ({yZ_Y(XX^\dagger)}+(1-y) {Z_\psi(XX^\dagger)}
\right )
XX^\dagger+\left ({Z_{YS}(XX^\dagger)\sqrt{y}}SX^\dagger+{\rm h.c.}
\right )
\label{wavefun}
\end{equation}
where $y=\gamma/(\gamma+2\beta)$.
Notice that the wave function mixing between $Y$ and $S$, which is radiatively
induced by the Yukawa couplings, leads to a corresponding term for $X$. 
The crucial point is that, while $Z_Y$ is only renormalized at 1-loop by matter
diagrams, $Z_\psi$ is also corrected by loops involving the massive $SU(2)_B$
vector superfields. By defining the Yukawa couplings at a renormalization
scale $\mur$ at which $Z_\psi=Z_Y=1$, the first term
in brackets in eq. \ref{wavefun} reads at lowest
order in $\ln(X/\mur)$
\begin{equation}
K \simeq XX^\dagger \left ( 1 - 
{(25\alpha^2 +4\beta^2)y+(\gamma^2 
- C_Bg_B^2)(1-y) \over 16\pi^2}\ln {XX^\dagger \over \mur^2}
 \right ) 
\label{approx}
\end{equation}
where $C_B$ is  the Casimir of $\psi$, equalling $4/3$ for $SU(2)_B$.
Now let us discuss supersymmetry breaking.
It occurs due to the non-perturbative $SU(5)_S$ dynamics which generates a 
superpotential linear in $X$. The source for this effect is
gaugino condensation in $SU(5)_S$ which for $X >> \Lambda$ behaves as
pure super-Yang-Mills theory.  
Gaugino condensation induces the non-perturbative contribution to the 
superpotential \cite{tvy,nsvz}
\begin{equation}
W = \langle \bar{\lambda}\lambda \rangle = \Lambda_L^3
\end{equation}
where $\Lambda_L$ is the low energy scale of the $SU(5)_S$ theory. This scale
is $X$ dependent, because so are the masses of the quarks we have integrated out.
Matching the low and the high energy gauge couplings\cite{russians} we get the 
effective superpotential for $X$
\begin{equation}
W = \alpha Y\Lambda^2=\rho X\Lambda^2
\end{equation}
where $\rho=\alpha\sqrt y$.
This superpotential breaks supersymmetry spontaneously: $F_X =
\rho\Lambda^2$. After taking into account 
the K\"ahler renormalization the effective potential for $X$ at one loop reads
\begin{equation}
V = K_X^{X-1}W_XW^{X*} ={\rho^2\Lambda^4\over y Z_Y(XX^\dagger)+
(1-y)Z_\psi(XX^\dagger)}
\end{equation}
The balancing between gauge and Yukawa coupling contributions in $Z_{Y}$, $Z_\psi$
can create local minima in $V$. This can be seen explicitly by using the
leading terms in eq. \ref{approx}. For a range of parameters, as it was the
case in ref. \cite{us}, the local minimum is at $X\gg \Lambda$, where our 
perturbative approximation to $K$ is reliable.
In this minimum the messengers have masses 
$\sim X$ with fermi-bose mass splitting due to 
$F_X\not = 0$. As a result, all the conditions of standard gauge-mediation
are  satisfied and supersymmetry breaking is transmitted to the MSSM
sector after integrating out $\bar Q, Q$. 

Now let us show how the $\mu$ and $B\mu$ terms are generated after 
supersymmetry 
breaking. This is arranged by the second term in 
(\ref{superpotential}) through the mechanism of \cite{dgp}.
Integrating the heavy mesons out and substituting
${\rm Tr}M = \Lambda^2$  we get the effective superpotential
\begin{equation}
W =  S(\lambda_1 H\bar H + \lambda_2 N^2/2  - \lambda \Lambda^2).
\label{weff}
\end{equation}
This has precisely the form introduced  in ref. \cite{dgp}. 
In the limit $Z_{YS}(\langle X\rangle)\sim 0$,
the field $X$ is decoupled (see eq. \ref{wavefun}) from the above sector 
in the effective theory below $\langle X\rangle$.
This situation is precisely the one assumed in ref. \cite{dgp} and we
will do the same in the following for the sake of simplicity. The 
conclusions are qualitatively unchanged in the general case of $Z_{YS}(
\langle X
\rangle)\sim 1$.
The dynamics that leads to nonzero values for $\mu$ and $B\mu$ can be
described in three steps.
\begin{description}
\item [a)] The superpotential in eq. \ref{weff},
induces at tree level the VEV for the $N$ field $N^2 = 
{2\lambda\over \lambda_2}\Lambda^2$. The other fields ($S, H, \bar H$) stay
zero to this order\footnote{For $S = 0$ there is a continuous degeneracy of the
vacuum parametrized by the holomorphic invariants $N^2$ and $H\bar{H}$ subject to 
the constraint $F_S = 0$. However, since at one-loop $S\neq 0$ (see the text 
below), the minimum is fixed at $H = \bar H = 0$
for a range of parameters \cite{dgp}.}
\item [b)] At one-loop a tadpole term for the scalar component of $S$ 
is induced \cite{dgp}
\begin{equation}
 -S{25 \lambda \rho^3 \over 16 \pi^2}\Lambda^4/\langle X \rangle + {\rm h.c.}
\end{equation}
(According to the rules of ref. \cite{gianriccardo}, this term can be seen
to arise from the renormalization of $Z_{YS}$).
This plays a crucial role, since it generates the VEV
$S = - {25 \rho^3 \over 32 \pi^2}\Lambda^2/\langle X \rangle $ and thus the 
$\mu$ term! A $B\mu$ term is also generated,
since for $S\neq 0$ the cancellation of $F_S$  is impossible and one
gets $B\mu = \lambda_1F_S = - {\lambda_1 \over \lambda_2}
\left ({25\rho^3 \over 32 \pi^2 } \right )^2\Lambda^4/\langle X \rangle^2$.  
\item [c)] At  two-loops there is an additional contribution to the
soft mass of $N$, which shifts the one-loop value of $B\mu$ so that
finally the total expressions for $\mu$ and $B\mu$ are \cite{dgp}
\begin{eqnarray}
\mu = {25 \rho^3\lambda_1 \over 32 \pi^2 \lambda_2}\Lambda^2/\langle X \rangle
\nonumber\\
  B\mu  = - {50 \lambda^2\lambda_1\lambda_2 \rho^2 \over
(16 \pi^2)^2}\left (1 + {25\rho^4 \over 8\lambda_2^2 \lambda^2} \right )
\Lambda^4/\langle X \rangle^2
\end{eqnarray}
\end{description}

Some remarks are now in order.
The first concerns the use of a singlet $Y$ rather that a non-singlet
$X$, as done in ref. \cite{us}, to give a mass to the messengers $Q\bar Q$. 
The reason for this choice
is that we want the same  invariant of the strong group to couple to both $X$
and $S$, otherwise a linear term in $S$ is not generated in $W_{eff}$.
 For instance one may try and use the $SU(6)^2$ model of ref. \cite{us}
and put $Y$ in the adjoint of $SU(6)_B$. In this case however $\langle M
\rangle $
is also in the adjoint, so that ${\rm Tr}\langle M\rangle =0$. The use 
of a singlet $Y$, and the addition of the $I,\psi,\bar\psi$ sector,
bypasses this difficulty.

The second remark concerns the genericity of our mechanism.
The superpotential in (1) is not the most general one could write down. 
However 
the results for $\mu$, $B \mu$  are qualitatively unchanged in a vast class of
models of this type. The only crucial feature of eq. (1)
which must be preserved is that $H$, $\bar H$ and $N$  appear
only via the expression $H\bar H+N^2$, as this is what allows $B\mu$ to be
a dynamical quantity. In the model we discussed, this property cannot be
enforced by any symmetry. However, as suggested in ref. \cite{dgp}, one 
can conceive a scenario where $N^2\to N\bar N$ so that $H\bar H+N\bar N$
is enforced by an  $SU(3)$ global symmetry. Such an effective symmetry
of the superpotential can indeed result as a subgroup of an original
$SU(6)$ GUT group in a class of models that naturally solve the doublet
triplet splitting problem \cite{su6}. With this assumption, eq. (1)
can be safely generalized by imposing an (anomalous) $R$ symmetry
under which the charges of $(Y,S,H\bar H,M,\phi^2,I)$ are $(2,2,0,0,4,-2)$.
Notice that $S$ and $Y$ have the same quantum numbers, but we can always
define $Y$ to be the combination that couples to $M$ and not to $H\bar H$.
Thus, in the most general renormalizable $W$ only the intermediate piece 
is modified to

\begin{equation}
I(\beta_1Y^2+\beta_2S^2+\beta_3YS-\gamma \psi\bar \psi).
\label{general}
\end{equation}
This preserves our mechanism: at $S=0$ there is still a flat direction $X$
along which supersymmetry is broken.
The only difference with respect to the previous case, is that
the field that gets a mass by pairing up with $I$ is now a linear combination
of $Y$, $\phi$ and $S$.

To conclude, notice that we have neglected the non-perturbative effects
of the balancing group. This is because the stabilization mechanism 
depends on a power of $g_B$ so that it works in the perturbative regime.
Non perturbative effects, on the other hand, go like ${\rm exp}-(8\pi^2/g_B^2)$
and become quickly negligible.  In our $SU(2)_B$  example there is an
instanton contribution to $W_{eff}\sim \Lambda_B^5/\phi^2$. This effect
restores supersymmetry somewhere along the $X$ line. However,
(as discussed in ref. \cite{us} in a similar case), it is consistent for us 
to neglect this term as long as it does not perturb the local mimimum
where we live. A direct estimate shows that this is the case already for 
$g_B(X)\leq 1$. Moreover one can imagine models where the balancing group
dynamics does not generate a superpotential. An example is $SU(2)_B$ with
matter content just given by a triplet $T$ with $\psi\bar \psi\to T^2$
in eq. (1).

\subsection{Generalizing the Vacuum Relaxation Mechanism for $\mu$}

The typical problem in GM theories is that $\mu$ and $B\mu$ are induced
at the same loop order resulting in
\begin{equation}
B\mu >> \mu^2.
\end{equation}
In the previous section we have shown a concrete GM model where a dynamical 
relaxation mechanism can avoid this problem. In this section we will generalize
 this mechanism and derive  sufficient conditions for it to work.
The key point is that both $\mu$ and
$B\mu$ depend on dynamical fields which, after supersymmetry breaking,
will adjust to the right vacuum expectation values due to
energy minimization. We note however, that all these additional fields
must have masses of order $F_X$, so that below this scale the low energy
spectrum is just that of the MSSM (plus possibly some additional light
states, like $X$, which only interact via 
$\langle X \rangle^{-1}$ suppressed couplings). This is the crucial difference
from the conventional singlet models \cite{DiNeSh} in which $\mu$ is induced as a
VEV of a {\it light} singlet field (an example of the later is the `sliding
singlet' mechanism for generating the $\mu$ term in GM theories\cite{alex}). 
The necessary ingredient of our approach is the
gauge-singlet field $S$ that couples to the Higgs doublets in the
superpotential (with coupling constant of order  one, which we do not display
for the time being). 
\begin{equation}
W = S H\bar{H}
\end{equation}
The expectation values of $S$ and $F_S$ will then respectively
set the scale of  $\mu$ and $B\mu$.
Following the notations of the previous section, let  $X$ be a superfield
that couples to the messengers ($\bar Q, Q$) and  breaks SUSY, through a
nonzero $F_X$ term. Then, to be in the right 
ball park the, the VEV of $S$ should be induced as a
one-loop effect from the scale $F_X \over \langle X \rangle$.
For this to be the case,
$S$ must couple at tree-level
to the messenger fields in the superpotential (just as in the
model of the previous section).
Now, the issue is to prevent $F_S$ from getting a VEV at the
same loop order. Our main remark is that, since $F_S$ contributes to the
 vacuum 
energy, it may be small just due to energetic reasons.
The sufficient condition for this to happen is that $F_S$ is a function of some
field(s)  which, at one-loop order,  has no potential apart from $F_S^*F_S$.
Thus the field(s) can  slide and compensate any source for $F_S$.
In our model such a field was $N$.
In this case $F_S$ and thus $B$ can only appear as higher loop-effects and 
$B\mu$ is acceptably small. Thus, the compensator field $N$ must have no
superpotential in the limit $S=0$ and we are lead to the following
general structure for the superpotential (again we omit the ${\cal O}(1)$ 
Yukawa
couplings)
\begin{equation}
 W = S ( H\bar{H} + f(N) + \bar QQ) + X\bar QQ
\end{equation}
In the model of the previous section we had $f(N) = \lambda_2N^2/2 - 
\lambda\Lambda^2$, and the scale $\Lambda$ came from the condensate of 
strongly coupled messengers $\bar QQ$.
Now we will study what is the possible general form of the $f$- function.
For the time
being we will be assuming a single fundamental scale $F_X \sim X^2$,
although this is not in any respect necessary for our mechanism to work
as it was clear from our model in which $X^2 >> F_X$.
The key idea is most transparent in units $\langle F_X \rangle = 1$.
In these units, as we will now show, the
function $f$ must satisfy the following  necessary conditions
in the minimum to zeroth-loop order:
\begin{equation}
f = 0,~~~ f' \sim 1,~~~f'' \sim 1 \label{fcond}
\end{equation}
where prime denotes derivative with respect to the compensator field $N$.
Now since $S$ is the only field that gets a correction to the potential from 
one-loop one-particle-irreducible diagrams, the effective
potential to this order can be written as (both Higgs doublets and messengers
 are put to zero as it should be)
\begin{equation}
V = |f|^2 + |S|^2|f'|^2 + V_1
\end{equation}
where $V_1$ is a one-loop effective potential for $S$. Since the $S$-VEV in 
any case will be stabilized by the second term, which gives a curvature
 $\sim 1$, it is sufficient to keep only a tadpole part linear in $S$ in
 $V_1$, which is of order $\epsilon S$.  In what follows
$\epsilon$  denotes a one-loop suppression factor.

Now, we  minimize with respect to $S$ and $N$  remembering that
$B\mu = F_s = f$, up to factors of order one, and
we get the following equations
\begin{equation}
  f^{*'}B\mu + f^{*''}f'|S|^2 = 0,~~~|f'|^2s + \epsilon = 0
\end{equation}
since the derivatives of $f$ are $\sim 1$ in the zeroth order, the same should
hold at one-loop, and, thus, we immediately get that
$\mu \sim S \sim \epsilon$ and $B\mu \sim \epsilon^2$. 

 The simplest explicit model with dynamical relaxation \cite{dgp} is based on
$f= N^2 - \Lambda^2$ and we have shown in the previous section  the scale 
$\Lambda$ can be indeed
generated dynamically by the condensate of the strongly coupled
messenger mesons.
What if the messengers are not transforming under a strong gauge group?  
It is interesting that in this case, the scale $\Lambda$ in $f$ can be induced
through the kinetic mixing of $X$ and $S$ superfields in the K\"ahler 
potential.
This the analog of the term $Z_{YS}$ in eq. \ref{wavefun}.
 Such a mixed term is not forbidden by any symmetry and
even if not present at the tree-level will be induced through the loops
with $Q$-particles. Assuming $Z_{XS}$ to be zero at the Planck scale $M_P$,
the resulting mixed term in the K\"ahler potential
has the form
\begin{equation}
 \Delta K \sim  {n5\lambda\rho \over 16\pi^2}{\rm ln}\left ({M_P^2 \over XX^*}
\right ) SX^*
\end{equation}
where $n$ is the number of messengers coupled to $S$.
After substituting $F_X \neq 0$
and solving the equation of motion of $F_S$, the effect of this term is 
just a
shift in $F_S$
\begin{equation}
  F_S\rightarrow \lambda_2N^2/2 +  F_X{n5\lambda\rho \over 16\pi^2}{\rm ln}
\left 
({M_P^2 \over XX^*}\right )
\end{equation}
The resulting $\mu \sim \sqrt{B\mu}$ term that is generated through the vacuum
relaxation
mechanism is suppressed by $1/\ln(M_P/X)$ factor with respect to the original
supersymmetry-breaking scale. For instance the ratio between the squark
mass and $\mu$ reads 
\begin{equation}
{\mu\over m_{\tilde Q}}=\sqrt{3\over 8n}{\lambda_1 \pi\over \lambda_2\alpha_3
\ln(M_P/X)}
\end{equation}
so that there seems to be a range of $n$ and $X$ where this ratio is close
to 1.

Finally we will mention one more logical possibility which can dynamically
introduce a scale of order $F_X$ in the function $f$ and satisfy conditions
(\ref{fcond}) necessary for the dynamical relaxation mechanism.
This is to have a SUSY breaking
sector communicating with the compensators through the $D$-term of some gauge
$U(1)$. Assume that the supersymmetry breaking VEVs develop along 
a $D$-non-flat direction of some $U(1)$ gauge factor.
This induces a non-zero expectation value of the $D$-term
$D \sim  g$ (in $F_X$ units) where $g$ is a $U(1)$ gauge-coupling constant.
Let us introduce a pair of  compensator superfields
$N_-$ and $N_+$ with charges $- 1$ and $+ 1$ under the above $U(1)$.
The function $f$ can then be chosen  as
$f=N_-N_+$. In the zeroth-loop order the
$N$-fields have an $F$-flat potential and one of them, (say $N_-$) can slide 
and
compensate the $D$-term picking up a VEV of order one. The conditions
(\ref{fcond}) are then automatically satisfied. The $\mu$ term is induced
as a one-loop effect through the usual tadpole term, whereas $B\mu$
is zero in this order and is only induced by higher loop corrections.

 We would like to thank Gian Giudice, Alex Pomarol, Lisa Randall and Tom Taylor
for discussions. One of us (S.D.) would like to thank Scott Thomas for
valuable discussions on the dynamical relaxation mechanism.

\begin{enumerate}

\def\ijmp#1#2#3{{ Int. Jour. Mod. Phys. }{\bf #1~}(19#2)~#3}
\def\pl#1#2#3{{ Phys. Lett. }{\bf B#1~}(19#2)~#3}
\def\zp#1#2#3{{ Z. Phys. }{\bf C#1~}(19#2)~#3}
\def\prl#1#2#3{{ Phys. Rev. Lett. }{\bf #1~}(19#2)~#3}
\def\rmp#1#2#3{{ Rev. Mod. Phys. }{\bf #1~}(19#2)~#3}
\def\prep#1#2#3{{ Phys. Rep. }{\bf #1~}(19#2)~#3}
\def\pr#1#2#3{{ Phys. Rev. }{\bf D#1~}(19#2)~#3}
\def\np#1#2#3{{ Nucl. Phys. }{\bf B#1~}(19#2)~#3}
\def\mpl#1#2#3{{ Mod. Phys. Lett. }{\bf #1~}(19#2)~#3}
\def\arnps#1#2#3{{ Annu. Rev. Nucl. Part. Sci. }{\bf
#1~}(19#2)~#3}
\def\sjnp#1#2#3{{ Sov. J. Nucl. Phys. }{\bf #1~}(19#2)~#3}
\def\jetp#1#2#3{{ JETP Lett. }{\bf #1~}(19#2)~#3}
\def\app#1#2#3{{ Acta Phys. Polon. }{\bf #1~}(19#2)~#3}
\def\rnc#1#2#3{{ Riv. Nuovo Cim. }{\bf #1~}(19#2)~#3}
\def\ap#1#2#3{{ Ann. Phys. }{\bf #1~}(19#2)~#3}
\def\ptp#1#2#3{{ Prog. Theor. Phys. }{\bf #1~}(19#2)~#3}

\end{enumerate}

\end{document}